\documentclass[aps,pra,twocolumn,amsmath,amssymb,floats,floatfix,showpacs]{revtex4}

\usepackage{graphicx}
\usepackage{dcolumn}

\begin{document}


\title{Quantum Brownian motion for periodic coupling
to an Ohmic bath} 

\author{J. Piilo}
\affiliation{
Department of Physics, University of Turku, 
FI-20014 Turun yliopisto, Finland
}

\author{S. Maniscalco}
\affiliation{
Department of Physics, University of Turku, 
FI-20014 Turun yliopisto, Finland
}

\author{K.-A. Suominen}
\affiliation{
Department of Physics, University of Turku, 
FI-20014 Turun yliopisto, Finland
}

\date{\today}

\begin{abstract}
We show theoretically how  
the periodic coupling between an engineered reservoir and
a quantum Brownian particle leads to the formation of a dynamical steady state
which is characterized by an effective temperature above
the temperature of the environment.
The average steady state energy of the system has a higher value than expected from the environmental properties.
The system experiences repeatedly a non-Markovian behavior -- as a consequence the corresponding effective decay for long evolution times is always on average stronger than the Markovian one. We also highlight the consequences of the scheme to the 
Zeno--anti-Zeno crossover which depends, in addition to the 
periodicity $\tau$, also on the total evolution time of the system. 
\end{abstract}


\maketitle

\section{Introduction}

Fundamental research on open 
quantum systems has traditionally focused on interactions between the reduced system and its
natural environment~\cite{Weiss99a,Breuer02a}. Recent theoretical and experimental developments in engineering the properties of the environments
open a new avenue by creating artificial controlled reservoirs with whom the reduced system interacts~\cite{Zoller96a,Myatt00a,Turchette00a}.
These developments make possible, e.g.,~to observe directly the appearance
of decoherence~\cite{Myatt00a,Turchette00a} and to study
non-Markovian dynamics of open systems in 
new regimes~\cite{Maniscalco04b,Maniscalco06a}.
Moreover, reservoir engineering and environment mediated schemes
allow the indirect control of quantum 
systems~\cite{Zoller04a,Romano06a} -- the quantum control being
the essential ingredient in the development of quantum simulators.
Generally speaking, the importance of the open system studies stems from their 
central role in understanding the quantum-classical 
border~\cite{Zurek02a,Zurek03a,Giulini03a} and 
from the role of decoherence as an obstacle for creating quantum
information processors~\cite{Nielsen00a,Stenholm05a}. Reservoir engineering provides a so far largely unexplored way to approach these issues which have both fundamental and applicative  character.

We focus on the reservoir engineering schemes for
a paradigmatic open system model, 
i.e.~quantum
Brownian motion 
(QBM)~\cite{Weiss99a,Breuer02a,Feynman63a,Caldeira83a,Haake85a,Grabert88a,Hu92a}, 
which has
a wide variety of applications ranging
from quantum optics~\cite{Qo} and nuclear physics~\cite{Np} to 
chemistry~\cite{Ch}. An earlier study has revealed various types
of non-Markovian
dynamics of QBM for structured reservoirs~\cite{Maniscalco04b}, i.e.~time-independent
engineered reservoirs~\cite{NoteOnT}. 
Here, we concentrate on a scheme where
the system is forced repeatedly to experience non-Markovian
behavior due to the structured  reservoir.
We show that this corresponds
to the case when the engineered reservoir is periodically switched
off and on -- and can also be seen as an initial step towards
more sophisticated time-dependent reservoir engineering
techniques. 
We call this switch off-on scheme as shuttered reservoir
and consider the cases where the duration of each period
is on the non-Markovian time-scale~\cite{NoteOnNM}.

So far time-dependent reservoir engineering schemes
have been considered mostly for few-level 
systems~\cite{KurizkiPRL}.
Here, we give an example of the effects of periodic
system-environment interaction for a more
complicated quantum system, namely harmonic
oscillator. Our results can find 
applications, e.g.,~in the field of linear quantum amplifiers.
In that context, the switching off-on action corresponds
to shuttering the pumping lasers, which may lead to
the interesting effects such as the enhancement 
or the reduction of the gain of the amplified 
field~\cite{LA}.

It is worth noting that artificial engineered reservoirs
has been recently created for trapped ion system
by applying electric noise to the trap electrodes and by using laser
light~\cite{Myatt00a,Turchette00a}.
Thus, one can think of creating shuttered reservoirs by
shuttering the noise which is applied to trap electrodes
or by shuttering the laser light. 

The theoretical description of the system dynamics, that we consider, is based on the recursive use of the appropriate master equation
and allows analytical solutions to be found.
For a small number of shuttering periods, one can control the
appearance of the 
quantum Zeno (QZE) and anti-Zeno effects 
(AZE)~\cite{Zeno1,Zeno2,Zeno3} for the reduced system by  tailoring the properties of the environment~\cite{Maniscalco06a}.
In this paper we concentrate on the intermediate 
time dynamics (intermediate number of periods)
and on the steady state properties of the reduced system for long time
(large number of periods).
We show that for a large number of cycles, the system always reaches
a steady state which is not in thermal equilibrium with the environment
at temperature $T$ but can be described by an effective
temperature $T_{\rm eff}>T$ and by the corresponding thermal state.
The steady state value of the average system energy 
is given by the ratio of the effective time averaged diffusion and dissipation constants.

The scheme has also interesting fundamental consequences for the 
appearance of QZE and AZE. As was 
shown in Ref.~\cite{Maniscalco06a},
for a small number of cycles, one can control the appearance of
the QZE or AZE by changing the environment parameters.
In this paper we show that
for large number of cycles the system always experiences an AZE.
In other words, the decay of the system is stronger
than the Markovian one for large number of periods.
Thus, for short evolution times 
the system may experience QZE which for long times
turns to AZE. This demonstrates that the appearance of quantum Zeno or anti-Zeno
effects in QBM depends also on the total evolution time of the system
(or number of measurements) and demonstrates the richness of
Zeno dynamics in QBM when compared to more simple systems, 
e.g,~two-level atoms.

The paper is organized in the following way. Section \ref{Sec:Qbm}
introduces the quantum Brownian motion model for a harmonically bound particle
and introduces the basic properties and parameters for the structured reservoirs we use.
These form the framework for Sec.~\ref{Sec:NonProj}
in which the formal treatment of open system with shuttered reservoir
is carried out leading to the concept of recursive master equation.
The results for the heating function dynamics 
are presented in Sec.~\ref{Sec:Dynamics}
and the discussion concludes the paper in 
Sec.~\ref{Sec:Con}.

\section{Quantum Brownian motion}
\label{Sec:Qbm}

We consider a harmonic
oscillator linearly coupled with a reservoir modelled as an
infinite set of non-interacting 
oscillators~\cite{Weiss99a,Breuer02a,Feynman63a,Caldeira83a,Haake85a,Grabert88a,Hu92a}.
The dynamics of a damped harmonic oscillator is described, in the secular approximation,
by means of the following generalized master equation in the interaction picture
\cite{Intravaia03a,Maniscalco04b}
\begin{eqnarray}
&&\frac{ d \rho(t)}{d t}= \frac{\Delta(t) \!+\! \gamma (t)}{2}
\left[2 a \rho(t) a^{\dag}- a^{\dag} a \rho(t)  - \rho(t)
a^{\dag} a \right]
\nonumber \\
&& +\frac{\Delta(t) \!-\! \gamma (t)}{2} \left[2 a^{\dag} \rho(t)
a - a a^{\dag} \rho(t) - \rho(t) a a^{\dag}
 \right]. \nonumber \\
 \label{Eq:MQbm}
\end{eqnarray}
In this equation, $a$ and $a^{\dag}$ are the annihilation
and creation operators, and $ \rho(t)$ is the reduced density matrix of the
system harmonic oscillator. The non-Markovianity is characterized
by the time dependence of the coefficients
$\Delta(t)$ and $\gamma(t)$ appearing in the master 
equation -- these  are
known as diffusion and dissipation coefficients, respectively
\cite{Intravaia03a,Maniscalco04b}. 

The diffusion coefficient appearing in the master equation (\ref{Eq:MQbm}), to
second order in the dimensionless coupling constant $g$, can be written 
in the units of $\hbar$ as
\cite{Caldeira83a,Maniscalco04b}
\begin{eqnarray}
\Delta(t) &=& 2 g^2 k_{\rm B} T \frac{r^2}{1+r^2} \left\{ 1
- e^{-\omega_c t} \left[ \cos (\omega_0 t)\right.\right.
 \nonumber \\
&&
-  (1/r)  \left. \sin (\omega_0 t )\right] \big\},
\label{eq:deltaHT}
\end{eqnarray}
where the assumption of the high-temperature
reservoir, $\bar{n}=k_{\rm B}T/\omega_0\gg1$ has been used.
The dissipation coefficient $\gamma(t)$ can be written as
\begin{equation}
\gamma (t)\! \!=\!\! \frac{g^2 \omega_0 r^2}{r^2+1} \left\{
1\!-\! e^{- \omega_c t} \left[ \cos(\omega_0 t) \! + r
\sin( \omega_0 t ) \right] \right\}. \label{gammasecord}
\end{equation}
Above, $r=\omega_c/\omega_0$ is the ratio between the environment
cut-off frequency $\omega_c$ and the oscillator frequency
$\omega_0$,  $k_{\rm B}$
the Boltzmann constant, and $T$ the temperature of the environment. 

A commonly used environment for open quantum systems is described by an Ohmic
reservoir spectral density with the Lorentz-Drude cut-off~\cite{Weiss99a}
\begin{equation}
J(\omega)= \frac{2  \omega}{\pi} \
\frac{\omega_c^2}{\omega_c^2+\omega^2}. \label{Eq:SpecDen}
\end{equation}
The spectral distribution is given by
\begin{eqnarray}
I(\omega) &=& J(\omega) [n_e(\omega)+1/2] \nonumber \\
&=& \frac{ \omega}{\pi} \frac{\omega_c^2}{\omega_c^2+\omega^2}
\coth{(\omega/K T)} , 
\label{Eq:I}
\end{eqnarray}
where $n_e(\omega)$ is the occupation of the environment mode at frequency
$\omega$ and Eq.~(\ref{Eq:SpecDen}) has been used. For high $T$,
Eq.~(\ref{Eq:I}) becomes
\begin{equation}
I(\omega) = \frac{2 k_{\rm B} T}{\pi}
\frac{\omega_c^2}{\omega_c^2+\omega^2}.
\label{Eq:IHt}
\end{equation}

The central parameter $r=\omega_c/\omega_0$ describes how on-resonant
the oscillator is with the reservoir.  
When $r>1$, intensive part of the environment spectrum overlaps
with the oscillator frequency and the decay coefficients $\Delta(t)\pm\gamma(t)>0$
for all times. Consequently, the master equation is of 
Lindblad-type~\cite{Lindblad,Gorini}.   
When $r<1$, the most intense part of the environment spectrum lies in the small
frequency range and the on-resonant intensity  is small. The decay coefficients $\Delta(t)\pm\gamma(t)$
acquire temporarily negative values
and the master equation is of non-Lindblad-type \cite{Maniscalco04a}.

The full solution of the master equation
(\ref{Eq:MQbm}) can be found e.g.~in Refs.~\cite{Maniscalco04b,Intravaia03a}.
In what follows we  study the time evolution of the heating
function $\langle n (t) \rangle$ with $n = a^{\dag} a$ quantum number operator.
The dynamics of $\langle n (t) \rangle$ depends only on the
diffusion coefficient $\Delta(t)$ and on the classical damping
coefficient $\gamma(t)$ \cite{Intravaia03a}. Furthermore, the
quantum number operator $n$ belongs to a class of observables not
influenced by the secular approximation \cite{Intravaia03a,Grabert88a}.

The solution for the heating function, valid for all times and all initial
states, is given by \cite{Maniscalco04b,Maniscalco04a}
\begin{equation}
\langle n(t) \rangle = e^{-\Gamma(t)} \langle n(0)
\rangle + \frac{1}{2} \left(  e^{-\Gamma(t)}  - 1\right) +
\Delta_{\Gamma}(t), 
\label{Eq:GenN}
\end{equation}
where the quantities $\Delta_{\Gamma}(t)$ and $\Gamma(t)$
are defined in terms of the diffusion and dissipation
coefficients $\Delta(t)$ and $\gamma(t)$ respectively  as follows
\begin{eqnarray}
\Gamma(t)&=& 2\int_0^t \gamma(t_1)\:dt_1, \label{Gamma} 
\label{Eq:Gamma}\\
\Delta_{\Gamma}(t) &=& e^{-\Gamma(t)}\int_0^t
e^{\Gamma(t_1)}\Delta(t_1)dt_1 \label{DeltaGamma}.
\label{Eq:DG}
\end{eqnarray}
For short times the system experiences non-Markovian dynamics and 
for long times, after Markovian heating,
the system reaches a thermal steady state with its environment.
This typical dynamics is modified 
in the context of the scheme presented in the next section.

\section{Shuttered reservoir and recursive master equation}
\label{Sec:NonProj}

We are interested in the
reduced system dynamics when its engineered artificial reservoir
is switched off and on in a repeated manner with 
period $\tau$ -- for this purpose we use the term
shuttered reservoir.
At the end of each period  the engineered reservoir is switched off, and switching on
initiates a new time evolution period for the system. 
Consequently, each
switching off and on process will reset the correlations
between the system and the environment. 
For an experimental realization of the scheme, one can, e.g., uncouple the system from
the engineered bath in a periodic way. The duration of the free evolution period has to be chosen 
long enough to avoid correlations between different system-bath coupling periods, 
i.e.,~the duration of the free evolution has to exceed the bath correlation time.

By resetting the correlations between the system and the environment,
the system oscillator is forced
to experience non-Markovian dynamics
in a periodic way.  Consequently, the system
remains repeatedly in the non-Markovian regime.
Following this scheme, the solution for the reduced density matrix and the heating
function dynamics can be obtained by the recursive use of the master
equation given by Eq.~(\ref{Eq:MQbm}). In other words, the master equation
is solved for each evolution period by using as an initial state of the system the one
obtained at the end of the previous period. In the case we consider,
the free evolution between system--reservoir interaction periods
can be neglected in the calculations~\cite{Free}.
  
This recursive use of the master equation
(\ref{Eq:MQbm}) leads, assuming 
an initial Fock-state, to the following equation
for the density matrix of the reduced system:
\begin{eqnarray}
\frac{ d \rho(t)}{d t} &=&
\gamma_1(\tau) \left[ a \rho (t) a^{\dagger} - \frac{1}{2} a^{\dagger}a\rho(t)
- \frac{1}{2} \rho(t) a^{\dagger}a\right]
\nonumber \\
&+& 
\gamma_{-1}(\tau) \left[  a^{\dagger} \rho (t) a - \frac{1}{2}a a^{\dagger}\rho(t)
- \frac{1}{2} \rho(t) a a^{\dagger}\right]. \nonumber \\
&&
\label{Eq:RecME}
\end{eqnarray}
Here, time $t=m\tau$ where $m$ indicates
the number of the period.
The decay coefficients are written  as
\begin{equation}
\gamma_1(\tau) = \frac{1}{\tau} \int_0^{\tau} dt' [\Delta(t')+\gamma(t')],
\end{equation}
and 
\begin{equation}
\gamma_{-1}(\tau) = \frac{1}{\tau} \int_0^{\tau} dt' [\Delta(t')-\gamma(t')].
\end{equation}
It can be shown that the master equation
(\ref{Eq:RecME}) corresponds formally to the case
where the reduced system dynamics is modified 
by periodic non-selective measurements. Thus,
our shuttered reservoir and initial state are equivalent 
to performing periodically non-selective
measurements of the energy of the oscillator~\cite{NoteOnGen}.
For a formal study of non-selective measurements
and master equations, see Ref.~\cite{Facchi05a}.

We are mostly interested in the case where the oscillator is initially in its ground Fock-state $|n=0\rangle$
(such as laser cooled single trapped ion) though the results for the steady state
presented in the following are valid for all initial states.
For an initial ground state, the solution of the master equation~(\ref{Eq:RecME}), in terms of the quantum characteristic function
(QCF) $\chi$ reads
\begin{equation}
\chi_{\rm t} (\xi) = \exp\left\{ - \left[ \langle n(t)\rangle +\frac{1}{2}\right] |\xi|^2 \right\},
\label{Eq:Chi}
\end{equation}
where the heating function is given by
\begin{eqnarray}
\langle n (t=m\tau) \rangle &=& 
\left( \frac{\Delta_{\Gamma}(\tau)}{1-e^{-\Gamma(\tau)}}-\frac{1}{2} \right)
\left( 1- e^{-m\Gamma(\tau)} \right).\nonumber \\
\label{Eq:NDyn}
\end{eqnarray}
The solution given in Eq.~(\ref{Eq:Chi}) is obtained by using
certain algebraic properties of the superoperators of the corresponding
generalized master equation, for more details 
see Refs.~\cite{Maniscalco04b,Maniscalco04a,Intravaia03a}.
Moreover, the QCF of Eq.~(\ref{Eq:Chi}) 
corresponds to 
a thermal state
at all times 
(c.f.~Ref.~\cite{Barnett})
and its time-dependence is given by $\langle n (t)\rangle$.

Eq.~(\ref{Eq:NDyn}) reveals that 
there exists a steady state value  for the heating function, $\langle n \rangle_{s}$,
and its value can be easily obtained
by taking the limit
of number $m$ of periods going to infinity. We obtain
from Eq.~(\ref{Eq:NDyn})
\begin{equation}
\langle n \rangle_{s} =
\lim_{m\rightarrow\infty} \langle n (t) \rangle =
\frac{\Delta_{\Gamma}(\tau)}{1-e^{-\Gamma(\tau)}}-\frac{1}{2}.
\label{Eq:Na}
\end{equation}
It is important to notice that the steady state value depends
on the duration $\tau$ of the periods. Moreover,
since $\Delta$ and $\gamma$ depend on $r$, consequently also $\Gamma$ and $\Delta_{\Gamma}$
depend on $r$ [c.f.~Eqs.~(\ref{Eq:Gamma}) and (\ref{Eq:DG})],
and the steady state
value may also depend on the form of the environmental spectral density applied in each period, and not only on
its temperature $T$. 

As a cross check, it is easy to see that with increasing duration of the period $\tau$, we obtain correctly the result  without shuttering.
The denominator in Eq.~(\ref{Eq:Na}),
$1/\left(1-e^{-\Gamma(\tau)}\right)$, goes to unity with increasing $\tau$
and 
we obtain $\langle n \rangle_{s} =  \Delta_{\Gamma}(\tau)-1/2$.
This matches with the long time result without the shuttering 
which can be calculated by
taking the limit $t\rightarrow \infty$ in Eq.~(\ref{Eq:GenN}).

To gain more physical insight, we expand $\exp\left[-\Gamma(\tau)\right]$  which appears explicitly 
and implicitly 
in Eq.~(\ref{Eq:Na}), with respect
to $\Gamma$.
Keeping the terms to lowest order in $\Gamma$,
and neglecting the term $1/2$ since we consider high temperature case, 
we obtain
\begin{equation}
\langle n  \rangle_{s} = 
\frac{1}{2}\frac{\int_0^{\tau}dt'\Delta(t')}{\int_0^{\tau}dt'\gamma(t')}.
\label{Eq:Ns}
\end{equation}
Moreover, this equation is valid also in the Markovian limit since
with the Markovian values for $\Delta$ and $\gamma$
we obtain $\langle n \rangle_{s} = k_{\rm B}T/\omega_0$, which matches
the high temperature Markovian result~\cite{NoteOnDecay}. 

Equation (\ref{Eq:Ns}) demonstrates
that the ratio between the time averaged diffusion and dissipation
gives the steady state value of the average system energy
in the presence of shuttered reservoirs. Here,
the average is taken over a single period of duration
$\tau$. The system
is not in thermal equilibrium with its environment and has an effective temperature
$T_{{\rm eff}}$ which is different than the temperature
$T$ of the environment.
Moreover, the steady state average energy of the system can be controlled 
by the period duration $\tau$ and the environment
"on-resonance" parameter $r$.

The steady state of the system is a thermal state, see Eq.~(\ref{Eq:Chi}),
 with $ \langle n (t)  \rangle = \langle n  \rangle_{s}$.
The corresponding effective temperature can be written as
\begin{equation}
T_{\rm eff} = Ê\frac{\omega_0}{ k_{\rm B}} \langle n  \rangle_{s}.
\end{equation}
This differs from the environment temperature $T$
since $\langle n  \rangle_{s} \neq {\bar n}$.
Moreover, since the system reaches an effective thermal state, it
eventually fulfills the detailed balance condition with appropriate
modifications to the transition rates given by the shuttered reservoir.

We have presented above the formalism and the cross-checks for its validity.
In the following section we continue with the results for the heating
function dynamics from 
Eq.~(\ref{Eq:NDyn})
and the steady state values of the average system energy from Eq.~(\ref{Eq:Na}).

\section{The system dynamics}
\label{Sec:Dynamics}
\subsection{Short and intermediate time dynamics}
\label{Sec:IntT}

\begin{figure}[tb]
\centering
\includegraphics[scale=0.4]{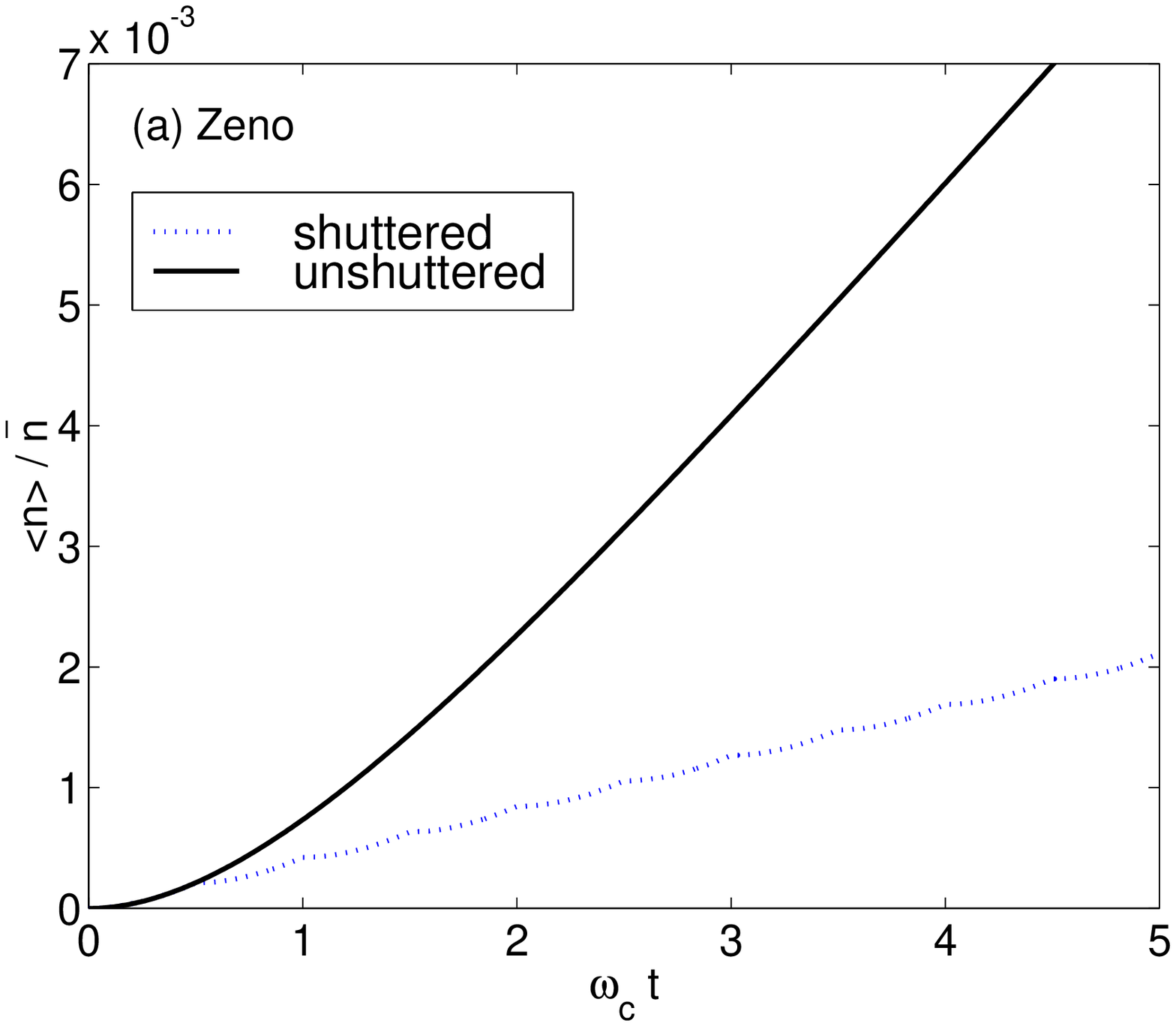}
\includegraphics[scale=0.4]{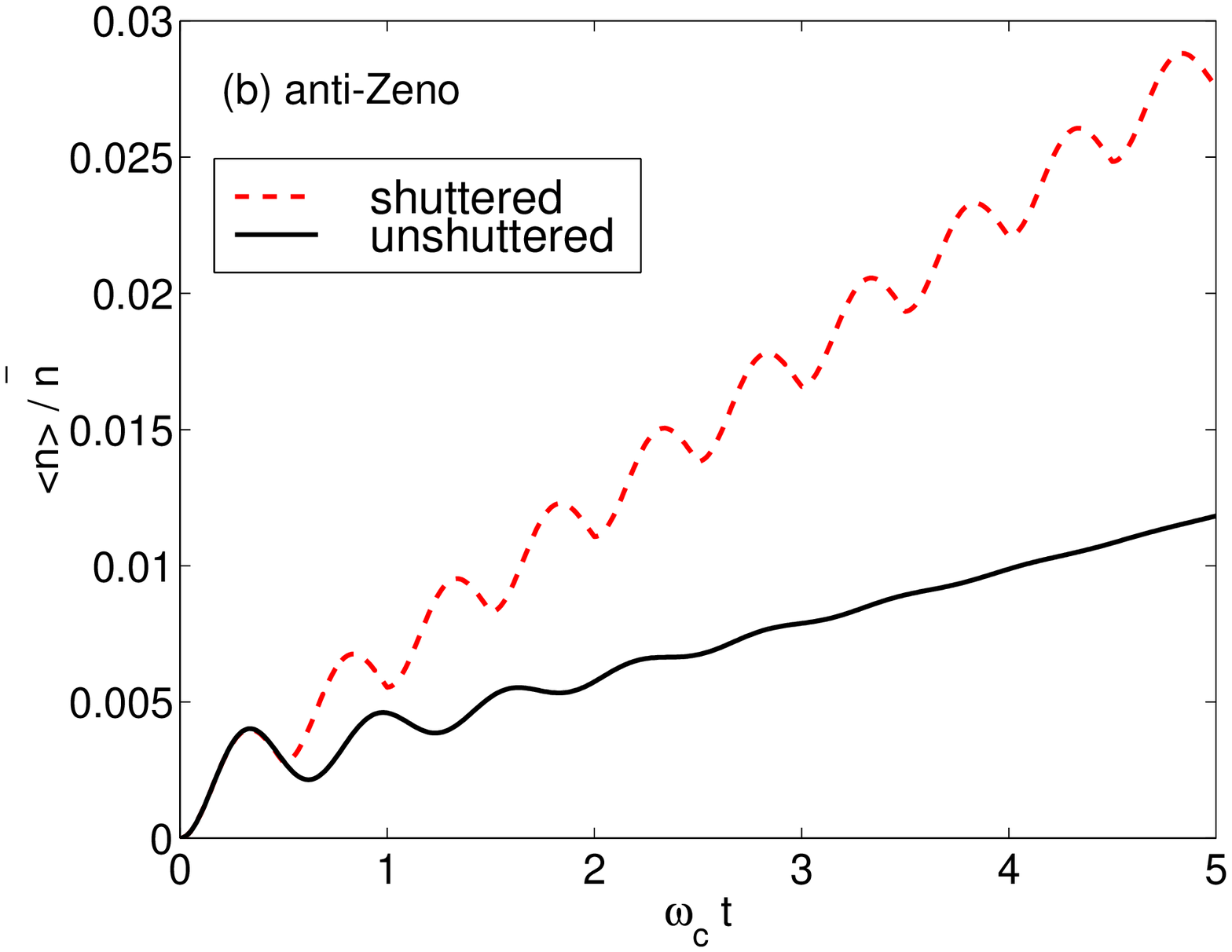}
\caption{\label{Fig:AntiZeno} 
(Color online) The short time behavior of the heating function displaying
(a) Zeno and (b) anti-Zeno effect.
In the Zeno effect, the shuttered reservoir reduces the heating of the oscillator while for the anti-Zeno case
the heating is enhanced when compared to the dynamics with the traditional unshuttered reservoir.
The parameters are $g=0.1$, $\tau=0.5 / \omega_c$, 
$\bar{n}=k_{\rm B}T/ \omega_0=10$. In (a) $r=10$ and (b) $r=0.1$.
}
\end{figure}

Figure \ref{Fig:AntiZeno}
displays the heating function dynamics over 
a couple of first shuttering periods.
It illustrates how the system exhibits (a) Zeno  and (b) anti-Zeno
effect. For various types of non-Markovian dynamics with traditional
unshuttered
reservoirs in this regime
see Fig.~1 in Ref.~\cite{Maniscalco04b} and for a study of the Zeno -- anti-Zeno crossover see Ref.~\cite{Maniscalco06a}.
The key aspect for this paper is that
both effects are possible for short evolution times.
Moreover,  one can control which one appears
by controlling the periodicity $\tau$ and 
the properties of the environment. Note that the figure
includes also the non-Markovian dynamics within each period
and not only the coarse graining in $\tau$ [c.f.~Eq.~(\ref{Eq:NDyn})].

\begin{figure}[tb]
\centering
\includegraphics[scale=0.4]{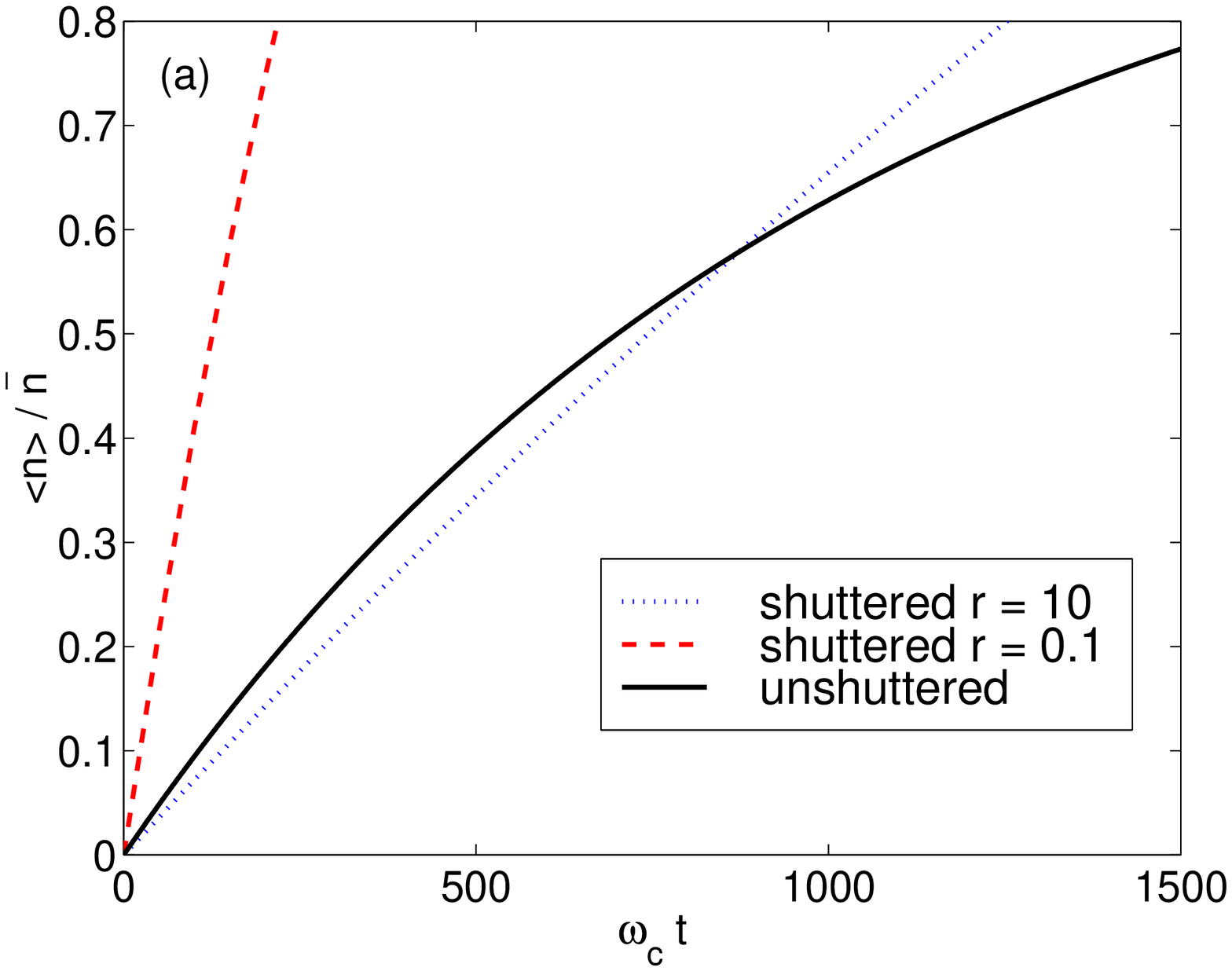}
\includegraphics[scale=0.4]{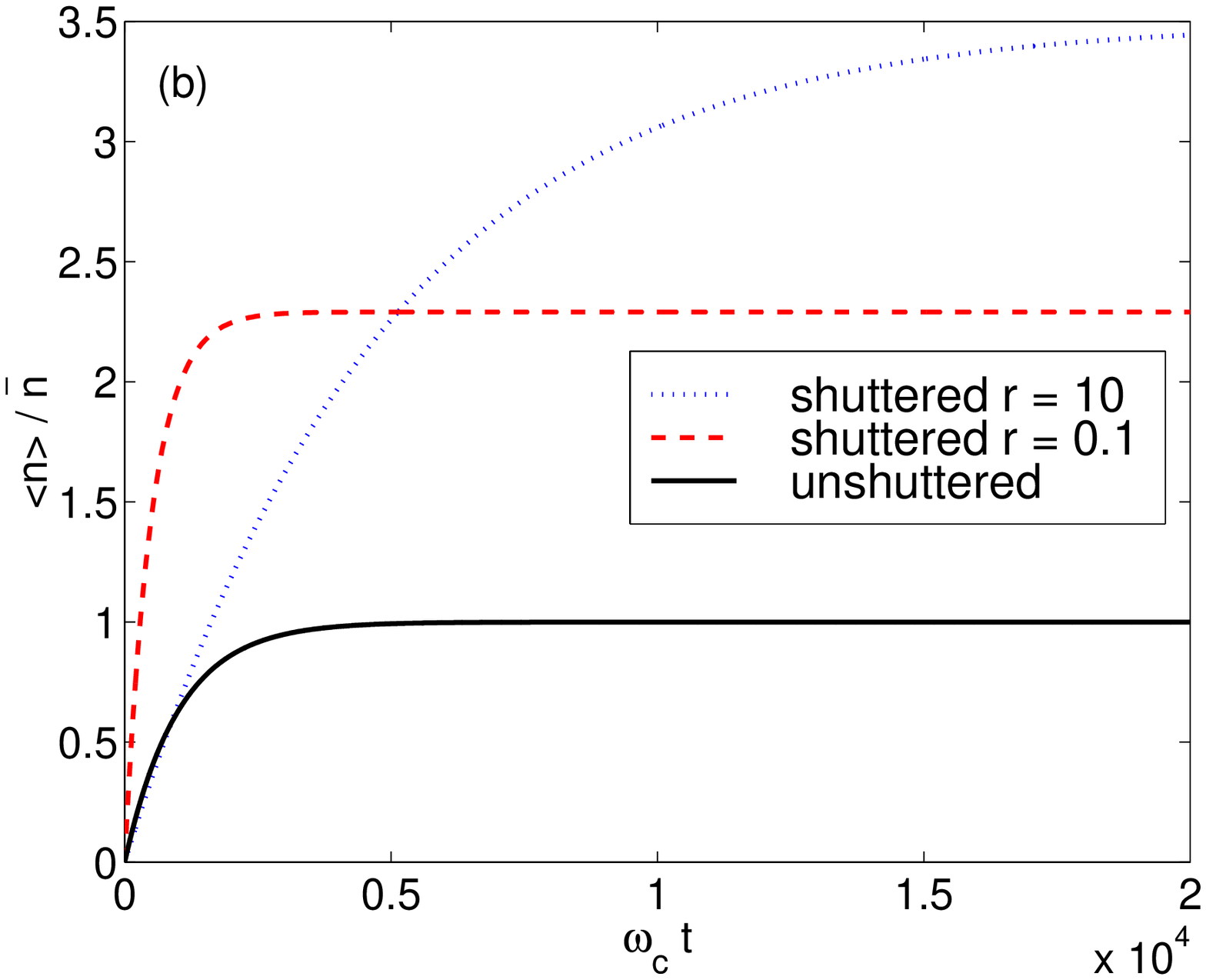}
\caption{\label{Fig:Ex1Ex2} 
(Color online) The heating function dynamics for intermediate times
when the system appproaches a steady state.
In panel (a) $0 \leqslant\omega_ct\leqslant 0.15\times10^4$ which shows the small time blow up of 
panel (b)
where $0 \leqslant\omega_ct\leqslant 10^4$.
The parameters correspond to those used in Fig.~\ref{Fig:AntiZeno}
except $\tau=1/\omega_c$. 
The solid line is for an unshuttered reservoir
showing the thermalization to 
$\langle n \rangle / \bar{n}=1$ in panel (b).
The dashed line is for the case of pure anti-Zeno dynamics by shuttered reservoir for all times.
The dotted line is for shuttered reservoir when
the dynamics begins as Zeno-type (reduced heating) but crosses later to anti-Zeno
behavior (increased heating). This crossing between the solid and dotted lines, which occurs 
at $\omega_ct\simeq 800$, is shown in detail in panel (a).
}
\end{figure}

Examples of the heating function dynamics for intermediate times, when the system approaches
the steady state, are displayed in Fig.~\ref{Fig:Ex1Ex2}.
The results demonstrate clearly that dynamics which begins with 
reduced heating, and therefore as Zeno-type
for short evolution times, becomes anti-Zeno for long times.
This can be observed as a crossing between the dotted 
and solid lines in Fig.~\ref{Fig:Ex1Ex2} (a) which shows
in detail the  passage from Zeno
to anti-Zeno region.

Moreover, Fig.~\ref{Fig:Ex1Ex2} (b) demonstrates that the approached steady state
value for the dotted line is around $3.5$
times the one expected from the temperature of the environment.
Furthermore, it is striking that there exists a crossing between the dotted and dashed lines
in Fig.~\ref{Fig:Ex1Ex2} (b). In another words, the dynamics which starts
as Zeno-type, shows for long times more pronounced anti-Zeno character than
the case which already initially begins as anti-Zeno type.
As a consequence, the highest steady state values of the energy
for long times are reached for the parameters that in short 
times give QZE and reduced decay.

The results above clearly demonstrate that the appearance of QZE and
AZE  is time-dependent in QBM. There exists a crossover between the two
effects in time. The fact that the appearance of these effects depends on the total
evolution time of the system, in addition of the periodicity $\tau$,  
is in contrast to more simple
systems like two-level atoms.
It has been shown in Ref.~\cite{Maniscalco06a}, which studied the controlling
of the Zeno and anti-Zeno
effects in QBM for short times, that the appearance
of the QZE and AZE depends on the quantum number $n$ for QBM.
Even though the time averages of the effective decay
coefficients remain fixed, this $n$-dependence
allows the crossover from Zeno to anti-Zeno effect to occur with time.
When the system heats, the expectation value $\langle n \rangle$ increases
with time, and the system experiences a crossover from Zeno
to anti-Zeno region.
This is not possible for a two-level system and demonstrates that
the Zeno--anti-Zeno phenomena 
has more variety in QBM than in more simple systems.

\subsection{Long-time dynamics and steady state}
\label{Sec:LongT}

The results presented in the previous subsection suggest that the heating function with shuttered reservoirs
approaches asymptotically
values which are higher 
than with the unshuttered reservoirs.
Actually, this is also indicated by the analytic solution
given in Eq.~(\ref{Eq:NDyn}) where the denominator is always
between $0$ and $1$.  
The system which is forced repeatedly to go through
the non-Markovian behavior, experiences enhanced dissipation compared to the 
Markovian one. The system reaches a steady state since the time dependent diffusion
and dissipation coefficients, $\Delta$ and $\gamma$ respectively, have
well defined time averaged values over $\tau$.

By using Eq.~(\ref{Eq:Na}) one can calculate the steady state
system energy as a function
of $\tau$, $r$, and the coupling constant $g$.
The magnitude of the coupling constant, in the region
of the validity of the model (weak coupling $g\ll1$) does not 
affect the steady state energy. 
Moreover, $\langle n \rangle_s$, as a function of $r$,
varies in a rather small range for $r\lesssim 1$ 
and then reaches a stable value for increasing $r$. Thus,
the most interesting parameter here
is the shuttering interval $\tau$. This conclusion
can be reached also by making a series
expansion of  $\langle n \rangle_s$, with respect
to small $\tau$. The dominant term contains
only $\tau$, and the parameters $r$ and $g$ come into play
only in the higher order terms.

\begin{figure}[tb]
\centering
\includegraphics[scale=0.4]{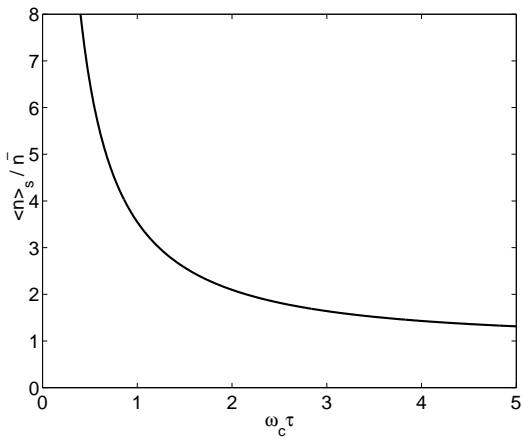}
\caption{\label{Fig:LimitTp} 
The steady state value of the heating function $\langle n \rangle_s$
as a function of the shuttering period $\tau$.
Parameters are $\alpha=0.1$, $r=10$, $\bar{n}=10$.
}
\end{figure}

Figure \ref{Fig:LimitTp} displays
the steady state $\langle n \rangle_s$ as a function
of $\tau$, where for other parameters some convenient
values has been chosen.
As expected, for increasing $\tau$, the average energy of the steady state 
approaches  the one with unshuttered reservoir and corresponds to the temperature of the environment.
When the duration of the periods $\tau$ is decreased, the steady state
energy of the system increases rapidly. This can be explained
by studying the behavior of $\Delta$ and $\gamma$ as a function of time.
For the typical spectral density we use in this paper
[c.f.~Eq.~(\ref{Eq:SpecDen})], $\Delta$ initially increases in the non-Markovian
region faster than $\gamma$
[c.f.~Eqs.~(\ref{eq:deltaHT}) and (\ref{gammasecord})].
The feature is enhanced in the time averaged values of these quantities
when $\tau$ decreases and demonstrates quantitatively the behavior
seen in Fig.~\ref{Fig:LimitTp}.

Naturally, the energy conservation has to be satisfied.
It requires work
to periodically shutter the reservoir. In other
words, it takes work to periodically couple the system
with the reservoir. This can be seen as the origin
of the increased system energy and
the energy is conserved since
the minimum work required exceeds the oscillator energy~\cite{OConnell}. 

\section{Discussion and conclusions}
\label{Sec:Con}

The creation of artificial environments and 
the reservoir engineering techniques open new avenues to study the non-Markovian
dynamics of open quantum systems and to improve the capability for 
the control of quantum systems. We have taken here an initial step towards time-dependent reservoir
engineering by considering the case when a structured reservoir is switched on and off in a periodic manner
(shuttered reservoir).

The results for quantum Brownian motion show
that the interaction of the reduced system with the shuttered reservoir changes the system
dynamics drastically compared to the conventional dynamics.
For short times the system may exhibit Zeno or anti--Zeno behavior
depending on the properties of the environment. A striking dynamical feature arises for
long-times -- the system always reaches a steady state in which the average system energy is larger than
the one corresponding to the the temperature of the environment. 
The shuttered reservoir forces
the system to experience the non-Markovian behavior periodically and the steady state
properties are consequently given by the time averaged properties of the
diffusion and dissipation between successive reservoir shuttering events.

One of the interesting consequences is that
the appearance of Zeno or anti-Zeno effects becomes dependent on the total evolution time of the system.
What starts as a Zeno effect for short and intermediate times turns to anti-Zeno effect
when the steady state regime is approached. The Zeno studies in simple systems, such as  a two-level system,
show how either Zeno or anti-Zeno effect appears. Here, we have demonstrated that 
a crossing between the two effects occurs with time and in
QBM the Zeno phenomena is very rich.

In the future, advanced reservoir engineering techniques may allow new ways to implement quantum
control. We have shown here that shuttered reservoir gives rise to interesting
dynamical effects. A next step is to study the case where the properties of the environment
remain the same for a given "on-period" but vary between different periods. 
Very precise ways to control the system properties by using reservoir engineering
combined with laser cooling may open new possibilities for the  development of general
quantum simulators for open systems whose dynamics is traditionally
tedious to solve by analytical means.

\acknowledgments

This work has been supported by the Academy of Finland 
(projects 108699, 115682, 115982) and the Magnus
Ehrnrooth Foundation.


\begin{references}
\bibitem{NoteOnT}
By a time independent reservoir we mean the ones whose spectral properties
remain unchanged with time.
\bibitem{NoteOnNM}
We have shown earlier that the non-Markovian time scale can be
prolonged to an experimentally accessible 
regime~\cite{Maniscalco04a}.
\bibitem{Free}
In our case, the state of the system is a thermal state
at all times. Since
a thermal state, being diagonal in the Fock-state basis,
is not affected by free-evolution, the periods
between two successive system-reservoir interaction periods can be neglected.
\bibitem{NoteOnDecay}
It is also possible to derive Eq.~(\ref{Eq:Ns}) by
considering the decay rates between the Fock-states
and assuming that in the steady state the average system
energy does not change in time.
\end{references}
\end{document}